\documentclass[preprint,showpacs,prl,aps]{revtex4}
\usepackage{amssymb}
\usepackage{amsmath}
\usepackage{amsfonts}
\usepackage{graphics}
\usepackage{epic}
\usepackage{eepic}
\usepackage{color}
\usepackage{epsfig}

\begin{document}

\def\ra{\rangle}
\def\la{\langle}
\def\bege{\begin{equation}}
\def\ende{\end{equation}}
\def\begarr{\begin{eqnarray}}
\def\endarr{\end{eqnarray}}
\def\ha{{\hat a}}
\def\hb{{\hat b}}
\def\hu{{\hat u}}
\def\hv{{\hat v}}
\def\hc{{\hat c}}
\def\hd{{\hat d}}
\def\no{\noindent}\def\non{\nonumber}
\def\hi{\hangindent=45pt}
\def\v{\vskip 12pt}

\newcommand{\bra}[1]{\left\langle #1 \right\vert}
\newcommand{\ket}[1]{\left\vert #1 \right\rangle}

\voffset=-2cm

\title{Towards Linear Optical Quantum Computers
}

\author{Jonathan P.\ Dowling$^1$} 
\author{James D.\ Franson$^2$}
\author{Hwang Lee$^1$}
\author{Gerald J.\ Milburn$^3$} 

\affiliation{
$^1$Quantum Computing Technologies Group, Section 367,\\ 
Jet Propulsion Laboratory, MS 126-347 \\
California Institute of Technology,
 4800 Oak Grove Drive, CA~91109 \\
$^2$Johns Hopkins University,
Applied Physics Laboratory, Laurel, MD 20723 \\
$^3$Centre for Quantum Computer Technology,
University of Queensland, QLD 4072, Australia
}

\date{February 4, 2004}

\begin{abstract}
Scalable quantum computation with linear optics was 
considered to be impossible due to the lack of efficient two-qubit logic gates, 
despite its ease of implementation of one-qubit gates.
Two-qubit gates necessarily need a nonlinear interaction
between the two photons, and the efficiency of this
nonlinear interaction is typically very tiny 
in bulk materials.
However, we recently have shown that
this barrier can be circumvented with effective nonlinearities
produced by projective measurements, 
and with this work 
linear-optical quantum computing becomes a new possibility of 
scalable quantum computation.
We review several issues concerning its principles and requirements.
\end{abstract}

\maketitle

\section{principles}

\no
There are three key principles in 
the Knill, Laflamme, and Milburn (KLM) proposal \cite{knill01} 
for efficient and 
scaleable quantum information processing (QIP):

\begin{itemize}
\item conditional non linear gates for two photon states
\item teleportation to achieve efficiency
\item error correction to achieve scalability
\end{itemize}

Conditional non linear gates are based on the non-unitary state change 
due to measurement. 
The gate works with some probability, 
but correct functioning is heralded by the measurement result. 
We seek to implement a nonlinear transformation (NS gate) 
on an arbitrary two photon state of a single mode field:
\begin{equation}
|\psi\rangle=\alpha_0|0\rangle_1+\alpha_1|1\rangle_1+
\alpha_2|2\rangle_1
\rightarrow|\psi'\rangle=
\alpha_0|0\rangle_1+\alpha_1|1\rangle_1-\alpha_2|2\rangle_1 \;.
\end{equation}

\begin{figure}[b]
\centerline{\psfig{figure=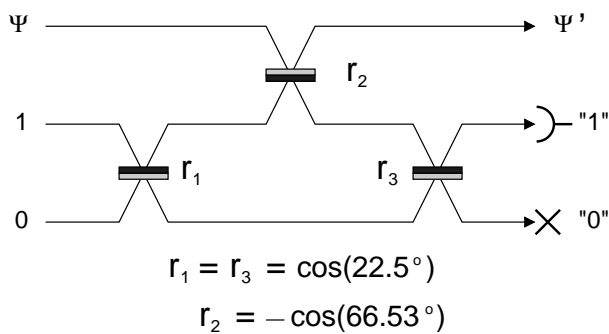,width=8cm,angle=0}}
\caption{
The conditional Nonlinear Sign (NS) gate.
}
\label{NS-gate}
\end{figure}

This is done using the linear optical network shown in figure \ref{NS-gate}. 
The signal state is first combined with two ancilla modes, 
one in a single photon state and one in the vacuum. 
At the end of the optical processing, photon counting is done on the ancilla modes. 
If the number of photons is unchanged from the input, 
the desired transformed state exits the signal mode port. 
This will happen with probability 0.25.

\begin{figure}[t]
\centerline{\psfig{figure=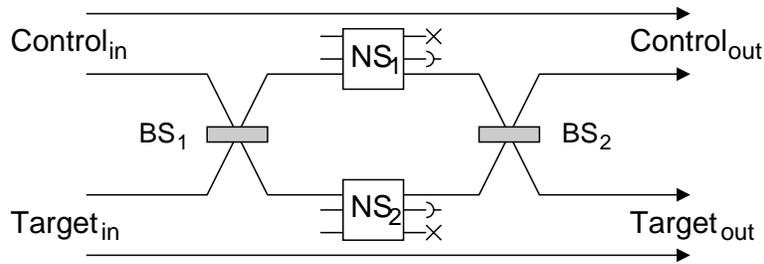,width=10cm,angle=0}}
\caption{
Controlled-$\sigma_z$ gate with dual-rail logic and two NS gates.
}
\label{c-sign}
\end{figure}

In order to use this result to implement QIP 
we  code the logical states as physical qubits using one photon in one of two modes: 
$|0\rangle_L  = |1\rangle_1\otimes|0\rangle_2 ,
\ \  |1\rangle_L  = |0\rangle_1\otimes|1\rangle_2$.
Single qubit gates are then implemented by a beam splitter.  
A two qubit gate, the conditional sign-flip gate, can then be implemented 
using the Hong-Ou-Mandel (HOM) interference effect 
to first convert two single modes each with one photon into 
an appropriate entangled two-photon state, figure \ref{c-sign}. 
Such a gate uses two NS gates and thus succeeds with probability $0.125$. 
A general formalism for the effective photon nonlinearities generated by
such conditional measurement schemes in linear optics 
can be found in Ref.~\cite{lapaire03}.

\begin{figure}[b]
\centerline{\psfig{figure=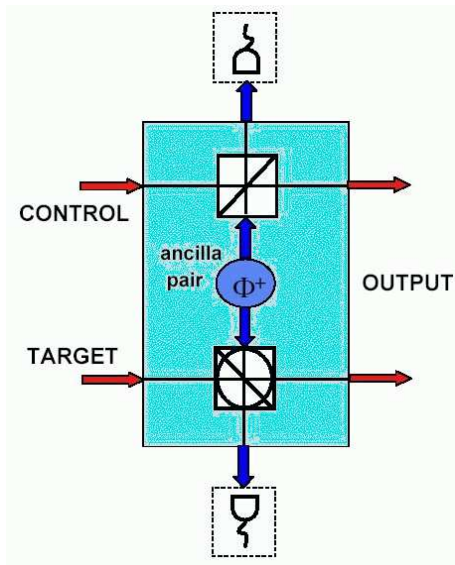,width=6cm,angle=0}}
\caption{
Implementation of a probabilistic controlled-NOT gate using polarization-encoded qubits.
}
\label{pol-cnot}
\end{figure}

We have also shown that probabilistic quantum logic operations 
can be performed 
using polarization-encoded qubits \cite{pittman01}, 
as illustrated by the controlled-NOT gate shown in Fig.~\ref{pol-cnot}.  
This device consists of two polarizing beam splitters 
and two polarization-sensitive detectors, 
along with a pair of entangled ancilla photons.  

The correct controlled-NOT logic operation will have been performed 
whenever one and only one photon is detected in each of the two detectors, 
which occurs with a probability of 0.25.  
Feed-forward control \cite{pittman02a} must also be applied, 
depending on what polarization states were measured.  
From an experimental perspective, 
this approach has the advantage of being relatively simple and insensitive to phase drifts.

A sequence of probabilistic gates is of course not scalable.
However, the Gottesman and Chuang protocol 
for implementing gates via teleportation \cite{gottesman99} 
can be used to fix this. 
Implementation of the gate then reduces to preparing the appropriate entangled state resource. 
That can be done off line using conditional gates and only when success is achieved is the teleportation gate completed. 
Using a resource with $n$ photons in $2n$ modes 
this decreases the gate failure  probability as $n^{-1}$,
or even as $n^{-2}$ \cite{franson02}.
Gates can thus be implemented efficiently.

When a teleportation gate fails 
it does so by an making a measurement of an incoming qubit. 
This is always heralded and can be fixed using detected measurement codes. 
This enables the scheme to be scalable (i.e., fault tolerant) 
provided the error probability is less than 0.5, 
but at the expense of very complicated multi-mode 
entangled resource states for teleportation. 
The other major source of error is photon loss. 
In principle this can also be corrected using teleportation gates. 
Scalability requires that loss probability per gate be less than 0.01. 
Not detecting a photon is equivalent to loss.

\begin{figure}[b]
\centerline{\psfig{figure=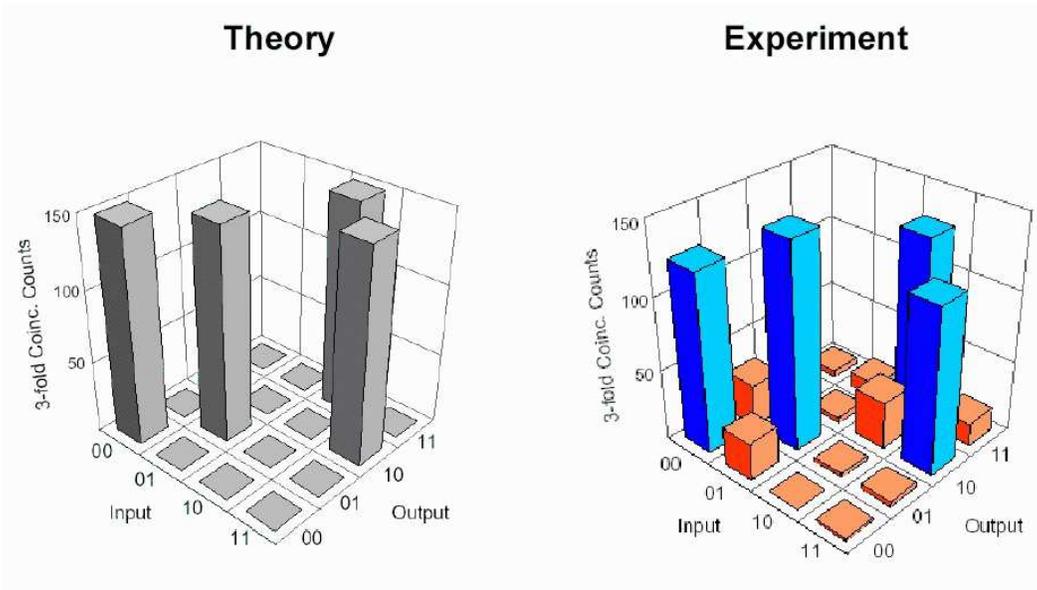,width=14cm,angle=0}}
\caption{
Implementation of a probabilistic CNOT gate 
using polarization-encoded qubits \cite{pittman03}.
}
\label{exp-cnot}
\end{figure}

Three experiments have implemented conditional two qubit gates:
Pittman et al. \cite{pittman03},
O'Brien et al. \cite{obrien03},
Sanaka et al. \cite{sanaka04}.
The first experiment \cite{pittman03} is based on the 
Pittman and Franson's polarization-encoded scheme. 
The second \cite{obrien03} is based on a simplification of the KLM-NS gate 
that requires only two photons. 
The last \cite{sanaka04} is a full four photon version of KLM. 
However, all experiments only work in the coincidence basis. 
This means that successful implementations correspond to 2 or 4 fold coincidence counts. 
However, no light leaves the device as all photons are detected.

Experimental results obtained from a CNOT gate in Ref.~\cite{pittman03} are shown 
in Fig.~\ref{exp-cnot}. 
Here a single ancilla photon was used, 
which restricts the operation of the device to the case in which 
a single photon is detected in each output port (the so-called coincidence basis). 
This was a three-photon experiment in which two of the single photons were obtained 
using parametric down-conversion while the third photon was obtained 
by attenuating the pump laser beam.  
Optical fibers were used instead of free-space components 
in order to reduce errors due to mode mismatch.  The fidelity of the output qubits was limited in this case 
by the degree of indistinguishability of the three photons.  
Experimental demonstrations of several other simple quantum logic gates 
have also been performed, including a quantum parity check \cite{pittman02}
and a quantum encoder \cite{pittman03b}.

\section{requirements}

\no
There are three major technical requirements that must be met:
\begin{itemize}
\item single photon sources
\item discriminating single photon detection
\item feed-forward control and  quantum memory
\end{itemize}

\no
The required ideal single photon sources are transform-limited pulses 
with one and only one photon per pulse. 
In practice this means that one must be able to exhibit HOM interference 
between photons from different pulses.
The required single photon detectors must be able to detect 
a single photon with efficiency greater than 0.99 and 
discriminate between 0, 1, and 2 photon counts \cite{achilles03}. 

\begin{figure}[t]
\centerline{\psfig{figure=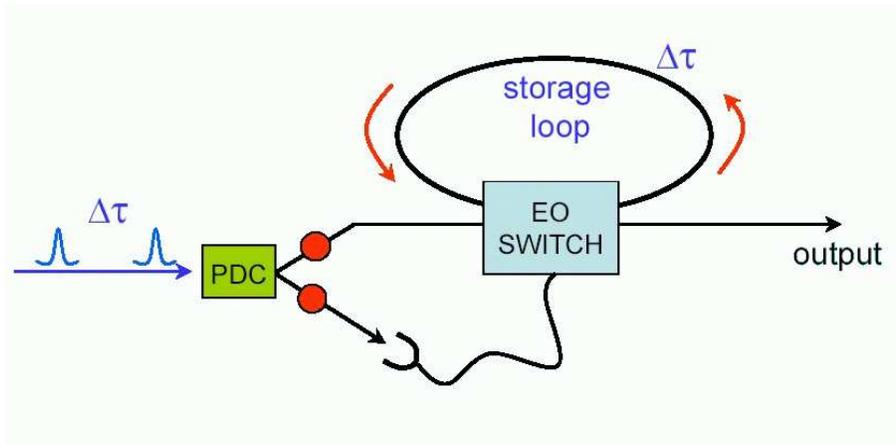,width=12cm,angle=0}}
\caption{
Single-photon source using parametric down-conversion 
and an optical storage loop.  
Similar storage techniques can also be used 
to implement a quantum memory device for single photons.
}
\label{pdc-spg}
\end{figure}

One approach to implementing such a single-photon source is illustrated 
in Fig.~\ref{pdc-spg}. 
A pulsed laser beam generates pairs of photons 
in a parametric down-conversion crystal.  
Detection of one member of a pair signals the presence of the other member of the pair, 
which is then switched into an optical storage loop.  
The single photon can then be 
switched out of the storage loop when needed \cite{pittman02b}.  
Although this kind of approach cannot produce photons at arbitrary times, 
it can produce them at periodic time intervals 
that could be synchronized with the cycle time of a quantum computer.  
A prototype experiment of this kind demonstrated 
the ability to store and retrieve single photons in this way, 
but its performance at the time was limited by losses in the optical switch.

The ability to switch a single photon into an optical storage loop 
and then retrieve it when needed can also be used 
to implement a quantum memory for single photons.  
This application is more demanding than the single-photon source described above, 
since the polarization state of the photons must be maintained 
in order to preserve the value of the qubits.  
A prototype experiment of this kind has also been performed, 
where the primary limitation was 
once again the losses in the optical switch \cite{pittman02c}.

\begin{figure}[b]
\centerline{\psfig{figure=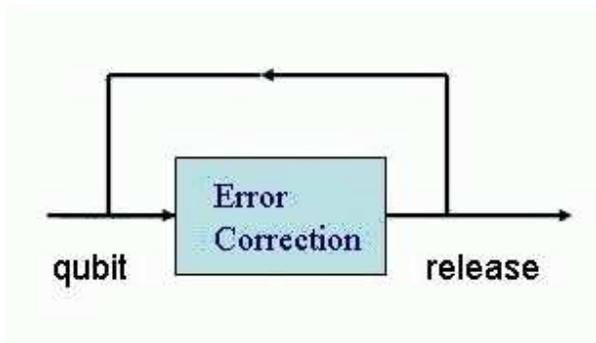,width=8cm,angle=0}}
\caption{
A cyclic quantum memory based on quantum error correction.
}
\label{ecc-box}
\end{figure}

\begin{figure}[b]
\centerline{\psfig{figure=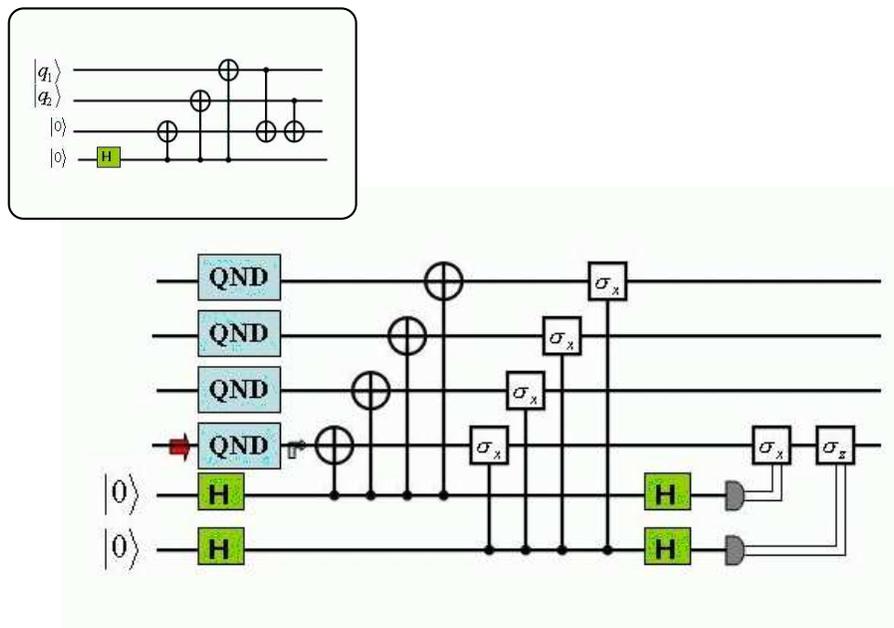,width=12cm,angle=0}}
\caption{
Quantum error-correction code that
recovers photon loss using two ancilla photons.
The QND box represents a single-photon quantum nondemolition measurement
device.
The inset shows the two-to-four qubit encoding.
}
\label{loqm-ecc}
\end{figure}

Furthermore, the ability to perform quantum logic operations using linear elements raises 
the possibility of using quantum error correction techniques 
to extend the coherent storage time of the quantum memory described above, 
see Fig.~\ref{ecc-box}.  
The primary source of error is expected to be photon loss, 
which can be corrected using a simple four-qubit encoding scheme \cite{gingrich03}
as illustrated in Fig.~\ref{loqm-ecc}.  
Provided that the errors in the logic gates and 
storage loops are sufficiently small, 
techniques of this kind can be used to store photonic qubits 
for an indefinitely long period of time.

An essential component in this kind of quantum memory
is the single-photon quantum nondemolition (QND) measurement device \cite{nogues99}.
Again a simple way to perform a single-photon QND measurement
can be provided by quantum teleportation technique.
If the input state is in a arbitrary superposition of 
zero and one photon with a fixed polarization,
the detector coincidence in Bell state measurement,
signals the present of a single
photon in the input and also the output states \cite{kok02}.

Similar techniques can also be used to compensate for the photon loss 
in optical fiber transmission lines, 
which would allow the development of a quantum repeater.
A quantum repeater is a device for 
achieving remote, shared entanglement by
using quantum purification
and swapping protocols \cite{briegel98}.
A simple protocol for optical quantum repeaters based on 
linear optical elements and an entangled-photon source
has been developed \cite{kok03}.
On the other hand,
utilizing quantum error correction, one can relay
an unknown quantum state with high fidelity down a quantum channel.
This device we call a quantum transponder,
and it has direct applications to quantum repeater and memory applications.

Quantum mechanics enables
exponentially more efficient algorithms 
than ones that can be implemented on a classical computer. 
This discovery has led to the explosive growth of 
the field of quantum computation  \cite{dowling03}. 
Many physical systems have been suggested for building a quantum computer,
but the final architecture is still to be determined.
These systems include ion traps, cavity QED, optical systems, quantum dots, 
nuclear magnetic resonance, and superconducting circuits.
In linear optical quantum computing, 
the desired nonlinearities come from projective measurements.

Projective measurements simply carry out measurements 
over some part of the quantum system, 
and project the rest of the system into a desired quantum state. 
Additional photons, known as ancilla, are mixed with the inputs 
to the logic devices using beam splitters while single-photon detectors 
are used to make measurements on the ancilla photons after the interaction. 
The nonlinear nature of single-photon detection 
and the quantum measurement process then project out the desired logical output. 
Therefore, although logic operations are inherently nonlinear, 
our approach uses only simple linear optical elements, 
such as beam splitters and phase shifters. 
Building a quantum computer will be a major challenge for 
a future quantum technology; 
requiring the ability to manipulate quantum-entangled states 
of large numbers of sub-components. 
Systematic development of each component of preparation, 
control, and measurements will facilitate the task of building a quantum computer.

\section*{Acknowledgements}
\no
Part of this work was carried out at the Jet Propulsion Laboratory, California
Institute of Technology, under a contract with the National Aeronautics 
and Space Administration. 
We wish to acknowledge support from 
the National Security Agency,
the Advanced Research and Development Activity, 
the Defense Advanced Research Projects Agency,
the National Reconnaissance Office, the Office of Naval Research,
the Army Research Office,
the IR\&D funding,
and the NASA Intelligent Systems Program.

\end{document}